\documentclass{mn2e}
\usepackage{psfig}
\usepackage{mnras_cite}
\newcommand{\redshifthi}{\left[ (1+z)/20 \right]}

\newcommand{\ESN}{\left( \frac{E_{VMS}}{10^{53} {\rm erg}} \right)}

\def\fun#1#2{\lower3.6pt\vbox{\baselineskip0pt\lineskip.9pt
  \ialign{$\mathsurround=0pt#1\hfil##\hfil$\crcr#2\crcr\sim\crcr}}}

\def\gtrsim{\mathrel{\mathpalette\fun >}}

\begin{document}

\title[Sunyaev-Zeldovich Fluctuations from the
  First Stars?]{Sunyaev-Zeldovich
  Fluctuations from the First Stars?}   
\author[Oh, Cooray \& Kamionkowski]{S. Peng Oh, Asantha Cooray \& Marc
  Kamionkowski\\
Theoretical Astrophysics, Mail Code 130-33, Caltech, Pasadena, CA
  91125, USA}

\maketitle
%%\pagerange{\pageref{firstpage}--\pageref{lastpage}} \pubyear{2002}
%%\label{firstpage}

\begin{abstract}
WMAP's detection of high electron-scattering optical depth
$\tau_{e}$ suggests substantial star formation at high redshift
$z \sim 17 \pm 5$. On the other hand, the
recovered $\sigma_{8} \sim 0.84 \pm 0.04$ disfavors a cluster
Sunyaev-Zeldovich (SZ) origin for the observed small-scale-CMB
fluctuation excess, which generally requires $\sigma_{8} \sim
1.1$. Here we consider the effects of high-redshift star
formation on the CMB. We derive a fairly model-independent relation
between $\tau_{e}$ and the number of ionizing photons emitted per
baryon $N_{\gamma}$, and use this to calibrate the
amount of high-redshift supernova activity.  The resulting
supernova remnants Compton cool against the CMB creating a
Compton-$y$ distortion $y \sim {\rm few} \times 10^{-6}$ 
within observational bounds. However they also
create small-scale SZ fluctuations, which could be comparable with
SZ fluctuations from unresolved galaxy clusters. This raises the
exciting possibility that we have already detected signatures of
the first stars not just once, but twice, in the CMB.
\end{abstract}

\section{Introduction}

The recent detection of high electron-scattering optical depth
$\tau = 0.17 \pm 0.04$ by the Wilkinson Microwave Anisotropy
Probe (WMAP) suggests a reionization redshift $z_{r}=17 \pm 5$
\cite{kogut,spergel}, providing good evidence for significant
star formation (SF) at high redshift $z$.  WMAP combined with other
large-scale structure data also supports a
$\Lambda$CDM cosmology with power-spectrum normalization
$\sigma_{8}=0.84 \pm 0.04$.

This power-spectrum normalization is discrepant from that
inferred from the CMB-fluctuation excess at small scales
\cite{mason,dawson}, if this excess is attributed to
the Sunyaev-Zeldovich (SZ) effect from unresolved groups and
clusters \cite{bond,komatsuseljak,goldstein}.  These observations require
$\sigma_{8} (\Omega_{b}h/0.035)^{0.29}=1.04 \pm 0.12$ at the 95\%
confidence level \cite{komatsuseljak}.

It has been argued that galactic winds could give rise to a detectable
SZ effect \cite{maj_nath}. Here we argue that the stellar activity
required to photoionize the Universe at $z_r\sim20$ injects a
considerable amount of energy into the IGM, which is then transferred
to the CMB due to the efficiency of Compton cooling at these high redshifts.
Although the resulting mean Compton-$y$ distortion is consistent
with the experimental upper limit, there may be detectable
angular fluctuations in the $y$ distortion.  We show, in fact,
that for reasonable reionization parameters the fluctuation
amplitude from high-$z$ SF may be comparable to
that from galaxy clusters.  If so, then the above-mentioned
discrepancy in the power-spectrum normalization may be resolved.

In the next Section we argue that supernova remnants at
$z\gtrsim10$ cool by Compton heating of the CMB and discuss the
energetics of this process.  In Section 3, we derive a relation
between the measured optical depth $\tau_e$ and the number of
ionizing photons required to reionize the Universe.  We then
show that this number of ionizing photons is proportional to the
energy injected into the IGM by supernovae, and thus the energy
transferred to the CMB.  In Section 4 we discuss angular
fluctuations in the $y$ distortion and show that they may be
comparable at small scales to those from unresolved clusters.

In all numerical estimates, we assume a $\Lambda$CDM cosmology given
by the best fits to the WMAP data: $(\Omega_{m}, \Omega_{\Lambda},
\Omega_{b}, h, \sigma_{8})=(0.27, 0.73, 0.044, 0.7, 0.84)$.  

\section{How does the Supernova Remnant Cool?}

At redshifts $z>7$, galactic winds powered by multiple
($>10^{5}$) supernovae (SN) or an energetic quasar jet are
cooled primarily by Compton cooling from the CMB
\cite{tegmark,voit,madau_ferrara}.  Less powerful winds result
in cooler remnants where radiative losses
could potentially be important. However, at $z\sim10-20$,
the wind from even a single SN will lose a substantial fraction of
its energy to the CMB, as we show below.

Zero-metallicity stars should be supermassive, ${M_{*}} \ge
100 {M_{\odot}}$, due to the
thermodynamics of molecular-hydrogen (${\rm H_{2}}$) cooling
\cite{abel,bromm02}. Furthermore, pair-instability
SN from such Very Massive Stars (VMSs) should have
explosion energies $\sim 100$ times more powerful than
conventional type II SN, $E_{\rm VMS} \sim 10^{53}$erg
\cite{heger_woosley}.  An extreme but plausible version of
zero-metallicity SF in low-mass halos $T_{vir} <
10^{4}$K prevalent at high $z$ is `one star per halo',
where internal UV photodissociation of ${\rm H_{2}}$ by the
first star in that halo halts all further gas
cooling and SF \cite{omukai,glover}. Simulations show the VMS quickly
photoevaporates all the gas within the
shallow halo potential well within a sound-crossing time
(M. Norman, private communication). In addition, the VMS
photoionizes a region around the halo up to $R \sim 70 ({\rm
M_{VMS}/100 M_{\odot}})^{1/3}$ kpc comoving, assuming that each
baryon in the VMS can ionize $\sim 10^{5}$ HI atoms
\cite{bromm01}. Thus, the SN remnant (SNR) expands into a
pre-ionized region at roughly the mean IGM gas density. During
the adiabatic Sedov phase, $R=\gamma_{o} (E
t^{2}/\rho_{IGM})^{1/5}$, where
$\gamma_{o}=1.17$. The remnant is no longer adiabatic and begins
to Compton cool when $t \approx t_{C}$, where the Compton
cooling time is,
\begin{equation}
     t_{C}=3 m_{e}c (4 \sigma_{T} a T_{\rm CMB}^{4})^{-1}= 1.4
     \times 10^{7} \redshifthi^{-4} {\rm yr},
\end{equation}
independent of temperature and density. The (proper) size of the remnant
at this point, when it quickly loses most of its energy, is
\begin{equation}
     R=2.2 \left( E_{VMS}/10^{53}\, {\rm erg} \right)^{1/5}
     \redshifthi^{-11/5} {\rm kpc}
\label{eqn:size_bubble}
\end{equation}
in physical units. The angular scale is $\theta=R/d_{A}=0.9^{``} \left({\rm
E_{VMS}}/{10^{53} {\rm erg}} \right)^{1/5} [(1+z)/20]^{-11/5}$ (which
corresponds to $l=\pi/\theta=7.6 \times 10^{5}$), beyond the reach
of present-day CMB interferometers. Thus, SNRs are effectively
point sources, unless many SN explode together in the same
galaxy, and/or SN bubbles from clustered halos overlap (see
below).

Most of the mass and energy of the remnant is in the dense
postshock shell, which is at $\rho_{\rm shell} \sim 4 \rho_{\rm
IGM}$. At $t=t_{C}$, we can compute the temperature behind the
shock front from the Sedov-Taylor solution, $v_{s}=0.4
\gamma_{o} (E/\rho_{\rm IGM}t^{3})^{1/5}$, and assuming a strong
shock $T_{c}=3 v_{s}^{2} \mu m_{p}/16 k_{B}$. We thus obtain the
ratio of Compton and isobaric radiative cooling time
$t_{rad}=2.5 k_{B}T/(n\Lambda(T))$ at $t=t_{C}$ as
\begin{equation}
     \frac{t_{rad}}{t_{C}}=0.4 \ESN^{0.4} \left(\frac{1+z}{20}
     \right)^{4.6} \Lambda_{23}^{-1},
\label{eqn:ratio_cooling}
\end{equation}
where $\Lambda(T)=\Lambda_{23}\, 10^{-23} {\rm erg \, s^{-1} cm^{3}}$,
and $\Lambda_{23}\sim1$ for low-metallicity gas with
$T\sim10^{5}-10^{7}$K. Thus, roughly a third of the SNR energy
is lost to Compton cooling.

The electron-ion equilibration timescale
\begin{equation}
t_{ei}=10^{5} \, {\rm yr} \left( \frac{1+z}{15} \right)^{-3} \left( \frac{\delta}{4}
\right)^{-1} \left( \frac{T}{10^{6} {\rm K}} \right)^{3/2} 
\end{equation}
(where $\delta$ is the overdensity of the postshock shell) is
significantly shorter than the Compton cooling time at all redshifts,
so there is no problem in quickly transferring the shock energy from protons
to electrons.  

A perhaps more likely scenario is one where many stars
$M_{*,tot} \sim 10^{7}
(f_{*}/0.1) (f_{b}/0.1) (M_{DM}/10^{9}) \,  M_{\odot}$ (where
$f_{b} \approx \Omega_{b}/\Omega_{m}$ is the baryon fraction,
and $f_{*}$ is the fraction of baryons which fragment to form
stars) form together in rarer, more massive halos $T_{vir}>
10^{4}$K where atomic cooling allows much higher gas densities
and more efficient SF \cite{oh_haiman}. The
massive-star evolution timescale is $t_{*} \sim 3 \times
10^{6}\,{\rm yr} \ll t_C$.  Thus, if SF takes place in a
starburst mode, the explosions are essentially simultaneous, and
$E_{tot} \approx N_{SN} E_{SN}$. Then, an extremely
energetic wind powers a much
hotter bubble, and from equation (\ref{eqn:ratio_cooling}),
$t_{rad}/t_{C} \propto E_{tot}^{0.4} \gg1$ and radiative cooling
is entirely negligible.  For instance, if
$f_{*} \sim 10\%$ of the baryons in a $M_{DM} \sim 10^{9}
M_{\odot}$ halo fragment to form VMSs, $t_{C} \sim t_{\rm rad}/40$). In
principle, radiative losses could be significant in the dense ISM of
these larger halos (since photoevaporation does not take place in
these deeper potential wells); however in practice most
simulations (e.g. \scite{maclow_ferrara}) find
that for such low-mass systems, the SN bubbles quickly `blow out'
(particularly in disks) and vent most of their energy and hot
gas into the surrounding IGM. Hereafter we shall encapsulate
this uncertainty as $\epsilon \approx 0.3-1$, the average
fraction of the explosion energy lost to the CMB via
Compton cooling. If stars form in clusters in higher-mass halos
rather than singly in low-mass halos we expect this efficiency
to be high, $\epsilon \ge 0.8$.
 
The spatial distribution of SF does not affect our
estimate of the mean Compton-$y$ distortion
($y=k_{B}T_{hot}/(m_{e}c^{2})\tau_{hot}$): more
clustered SF results in higher $T_{e}$ but lower
$\tau_{hot}$. However, it does of course affect the strength
of SZ fluctuations. We now turn to these issues.    

\section{Thermal Sunyaev-Zeldovich effects}

\subsection{SZ flux from Individual Supernovae}

The SZ flux from an individual SNR is
\begin{eqnarray}
     S_{\nu}&=&\frac{2k_{B}^{3}T_{\gamma}^{2}}{h^{2}c^{2}}g(x) \int d\Omega
     |\Delta T_{\nu} (\theta)| \\ \nonumber
     &=&
     \frac{2k_{B}^{3}T_{\gamma}^{3}}{h^{2}c^{2}}g(x)
     \frac{k_{B}T_{e}}{m_{e} c^{2}} \sigma_{T}
     \frac{N_{e}}{d_{A}^{2}} \\ \nonumber 
     &=&1.8 \times 10^{-2} \left(\frac{g(x)}{4} \right) \left(
     \frac{E_{VMS}}{10^{53}\, {\rm erg}} \right) \left( \frac{\epsilon}{0.5}
     \right) \left(\frac{z}{20}\right)^{2} {\rm nJy}
     \label{eqn:SZ_flux}
\end{eqnarray}
where $g(x)= x^{4}{\rm e}^{x} [x {\rm coth}(x/2)-4]/({\rm e}^{x}-1)^{2}$ is the spectral function, $x \equiv h\nu/kT_{\gamma}$,
$T_{\gamma}=2.7 {\rm K}$ is the CMB temperature, and $N_{e}$ is the total
number of hot electrons at temperature $T_{e}$. In the second line we have
used $k_{b} T_{e} N_{e} \approx E_{VMS}$.  The energy of the
remnant is a function of time, $E_{VMS}(t) \approx E_{\rm VMS,o}
{\rm exp}(-t/t_{C})$ (in the regime where Compton cooling off the
CMB dominates). The flux from an individual SNR is
well beyond threshold for any realistic experiment; only
a very large number of SN ($>10^{8}$) going off
simultaneously within a star cluster will be detectable at the
$\sim$ mJy level. Thus, SN bubbles cannot be identified and removed
from SZ maps; unresolved SN will create both a mean Compton-$y$
distortion and temperature fluctuations, which we now calculate.

\subsection{Mean Compton $y$ distortion}
\label{section:y_parameter}

We first use the observed optical depth $\tau_e$ to derive a lower
limit to the number of ionizing photons $N_\gamma$ emitted per
baryon. The dominant contribution to $\tau \propto
(1+z)^{1.5}$ comes from high $z$ where the recombination
time $t_{rec} \propto (1+z)^{-3}$ is short,
and recombinations are the rate-limiting step toward achieving
reionization. The filling factor of HII regions is
$Q_{HII} \approx t_{rec}/t_{ion} \approx \dot{N}_{\gamma}/(\alpha_{B}
n_{e}(z) C_{II}(z))$, where $\dot{N}_{\gamma}$ is the rate at
which ionizing photons are emitted per baryon (in
units of $s^{-1}$), $C_{II} \equiv \langle n_{e}^{2} \rangle/ \langle
n_{e} \rangle ^{2}$ is the clumping factor of
ionized regions (e.g., \scite{madauetal}). The clumping factor
increases with time as structure formation proceeds; it declines
sharply at high $z$ and is $C \approx 2$ at $z=20$, compared to
$C \approx 30$ at $z=10$ \cite{haimanabelmadau}. More sophisticated
considerations \cite{jordietal} take into account the density
dependence of reionization, but apply primarily near the epoch of
overlap, $Q_{II} \rightarrow 1$, when overdense regions are
ionized. This has little impact on our estimates. Most of the
mass and the optical depth comes from regions close to the mean
density.

The electron-scattering optical depth is given by:
\begin{equation}
     \tau_{e} = c \sigma_{T} \int dz \frac{dt}{dz} n_{e}(z) {\rm min} \left(1,
     \frac{\dot{N}_{\gamma}}{\alpha_{B} n_{e}(z) C_{II}} \right) = \frac{c
     \sigma_{T} N_{\gamma}}{\alpha_{B} C_{II}}.
\label{eqn:tau}
\end{equation}
Due to the cancellation of the electron density, this expression
is {\it independent} of the redshift of reionization, and the
evolution of the comoving
emissivity $\dot{N}_{\gamma}(z)$ with redshift; it allows us to directly
relate $\tau_{e}$ and $N_{\gamma}$. The only redshift dependence
lies in the effective clumping factor $C_{II}$, which increases
if reionization takes place at late times. The second equality
breaks down if overlap $Q_{II} \rightarrow 1$ is achieved at
high $z$ and $ \dot{N}_{\gamma}/(\alpha_{B} n_{e}(z)
C_{II}(z)) >1$ (i.e., recombinations no longer balance
ionizations); in using the expression we would then
underestimate $N_{\gamma}$, which would only imply an even larger
emissivity. The high optical depth $\tau_{e}=0.17 \pm 0.08$ ($2\sigma$)
\cite{kogut,spergel} detected by WMAP therefore implies that
\begin{equation}
     N_{\gamma}^{IGM}= 17 \pm 8 \left( \bar{T}/10^{4} \right)^{-0.7}
     \left(C_{II}/4 \right)  
\end{equation}
ionizing photons were emitted per baryon, where $\bar{T}$ is the
mass-weighted temperature of the reionized IGM (the $\bar{T}^{-0.7}$
factor arises from the temperature dependence of the recombination
coefficient). Consistency with WMAP requires {\rm more} SF if
reionization took place at lower redshift, due to the increase
in gas clumping at late times.

Since only a fraction $f_{\rm esc}$ of ionizing photons escape from their
host halo due to photoelectric absorption, the actual total number of
ionizing photons produced is larger, $N^{tot}_{\gamma}=
N^{IGM}_{\gamma} f_{\rm esc}^{-1}$. In addition, we only care about
those photons emitted at $z>6$, when $t_{C}
< t_{H}$ (where $t_{H}$ is the Hubble time) and Compton cooling is most efficient. Since $\tau_{e}(z<6)
\approx 0.05$, we have $\tau_{e}(z>6) \approx 0.12$; therefore
$0.12/0.17 \sim 0.7$ of the photons are emitted at $z>6$. Thus: 
\begin{equation}
     N^{tot}_{\gamma}(z>6) \approx 25
     \left(f_{\rm esc}/0.3\right)^{-1} \left( \tau_{e}(z>6)/0.12
     \right) \left(C_{II}/4 \right).
\label{eqn:Nphoton}
\end{equation} 
Estimates for the escape fraction span $f_{\rm esc} \sim
10^{-2}-1$, but if the earliest stars reside in
low-mass halos with $T_{vir} < 10^{4}$ K, the gas in such halos
is quickly photoionized and driven out in a photoevaporating
wind.  If so, $f_{\rm esc} \sim {\rm few} \times 0.1$ to $f_{\rm
esc} \sim 1$.
 
How much SF and energy production is associated with
$N_{\gamma}^{tot}$?  We consider first VMSs, supported as the
source of reionization perhaps by elemental-abundance evidence 
from low-metallicity halo stars
\cite{ohetal} and theoretical modelling \cite{cen,wyithe_loeb}. \scite{bromm01}
find that for $300 M_{\odot} < M_{*} < 1000
M_{*}$, the luminosity per solar mass is approximately constant; for
$M_{*} \sim 100 M_{\odot}$, it falls by a factor of 2. Our
estimates are thus independent of IMF details. For 1
ionizing photon per baryon in the universe, $f_{*} \sim 10^{-5}$
baryons have to be processed into VMSs; thus,
$N_{\gamma}^{tot}=25$ corresponds to $f_{*} \sim 2.5 \times
10^{-4}$. A $\sim 100 {M_{\odot}}$ pair-instability SN
releases $E_{VMS} \sim 10^{53}$erg \cite{heger_woosley}, or
$E_{b} \sim 0.5$MeV per baryon processed into the VMS.
The total energy release per baryon is therefore:
\begin{equation}
     E_{c}= \epsilon f_{*} E_{b} = 100 \left( \epsilon/0.8 \right)
     \left( N_{\gamma}^{tot}/25\right)\, {\rm eV},
\end{equation} 
where $\epsilon$ is the fraction of the thermal
energy which is lost to the CMB. A possible caveat is if a large fraction of the mass in
the first stars went into VMSs with $M_{*} > 260 M_{\odot}$,
which may collapse directly to black holes without
exploding as SN \cite{heger_woosley}.

The fraction of baryons processed into VMSs $f_{*}
\sim 2.5 \times 10^{-4} (N_{\gamma}/25)$ implies an IGM metallicity $Z
\sim 6 \times 10^{-3} Z_{\odot}$, assuming uniform enrichment (since
$\sim$half the VMS mass is thought to end up as metals).  This is
consistent with the observed metallicity of the Ly$\alpha$ forest at
$z=3$ of $Z \approx 10^{-2.5} Z_{\odot}$, which is not observed
to evolve strongly at higher $z$ \cite{song01}. Thus, the
metals seen in the Ly$\alpha$ forest may well have been
injected at very high $z$ by Pop III stars. No trace of the
entropy injection associated with the metal-polluting winds would
remain, due to the high efficiency of Compton cooling.  

Our derived ionizing-photon:energy:metal ratios would also hold
for normal stellar populations (rather than VMSs), which produce
roughly the same amount of SN energy and metals per ionizing photon.  The arguments are also roughly independent of IMF, as
the massive stars that emit ionizing photons also eventually
explode as SN.

We now compute the Compton-$y$ parameter associated with this
energy injection. For simplicity, we assume that all of the
energy is injected at some redshift $z_{i}$. The actual redshift
evolution introduces at most a factor $\sim 2$ uncertainty (see
expresion below). The $y$
parameter is then given by
\begin{eqnarray}
     y&=&\frac{c \sigma_{T}}{m_{e} c^{2}} \int_{t_{i}}^{t_{o}} dt\,
     n_{e}(t)E_{c,o} e^{-(t-t_{i})/t_{C}}\\ \nonumber   
     & \approx&  n_{e}(z_{i}) \sigma_{T} c t_{C}(z_{i})
     \frac{E_{c}}{m_{e} c^{2}} \\ \nonumber
     &=& 3.6 \times 10^{-6} \left( (1+z_{i}/15\right)^{-1} \left( 
     E_{c}/100\, {\rm eV}\right),
\label{eqn:y_parameter}
\end{eqnarray} 
where we have moved the electron density outside the integrand,
$n_{e}(t) \approx$const, since the density does not change
significantly on the timescale over which the gas Compton cools.
In the RJ limit, $(\Delta T/T)=-2y=7 \times 
10^{-6}$.  The $y$ distortion is less than the COBE FIRAS
constraint, $y \le 1.5 \times 10^{-5}$ \cite{fixen}, as it
should be. Such a $y$ distortion could in principle be detected by
future instruments \cite{fixsen_mather}. In addition, a low-frequency
distortion due to free-free emission from ionized halos should also be
detectable \cite{oh99}.  

We pause here for a simple order-of-magnitude check.  Let the
total amount of energy per baryon injected through Compton
cooling into the CMB be $E_{c}$. If this takes place at some
median redshift $z_{i}$, this introduces an energy density
perturbation of the CMB $\Delta U_{\gamma} \sim n_{b} E_{c} \sim
6.8 \times 10^{-2} ([1+z]/15)^{3} (E_{c}/100 \, {\rm eV}) {\rm
eV \, cm^{-3}}$. The CMB energy density is $U_{\gamma}= 1.3
\times 10^{4} [(1+z)/15]^{4} {\rm eV \, cm^{-3}}$ resulting
in a temperature perturbation,
\begin{equation}
     \frac{\Delta T}{T_{\gamma}} \sim \frac{1}{4} \frac{\Delta U}{U}
     \sim 5.2 \times 10^{-6} \left( \frac{1+z}{15} \right)^{-1} \left(
     \frac{E_{c}}{100 \, {\rm eV}} \right) 
\end{equation}
roughly consistent with our previous estimate, from $(\Delta T/T
)=-2y$. Why is the mean $y$ distortion
due to non-gravitational heating by high-$z$ SN competitive with
that from galaxy clusters today?  By integrating over the
Press-Schechter mass function and assuming
$T_{gas}=T_{vir}$, we find that the mean mass-weighted
gas temperature today is $\langle T \rangle =0.7$ keV.  However
the Compton cooling time in clusters is $t_{C} \sim 150
t_{H}$, so only $\epsilon \sim 1/150$ of that energy is
extracted. Since $y \propto E_{c} (1+z)^{-1}$, we find that
the $y$ distortion due to clusters is $\sim (0.7
{\rm keV}/150)/{\rm 0.1 keV} \times 15 \sim 1$ times the
distortion due to high-$z$ SN.

\section{SZ fluctuations}

Angular SZ fluctuations can be induced by Poisson fluctuations
in the number density of sources, as well as by clustering of
the underlying mass distribution.  Poisson fluctuations are the
dominant source of SZ fluctuations for galaxy clusters, but they
are negligibly small for high-$z$ SN for the following reason:
like high-$z$ halos, clusters are $\sim 2-3 \sigma$ fluctuations
at the epoch at which they form and contain roughly the same
fraction of collapsed mass; however, they are more massive by
$\sim 6$ orders of magnitude and hence have a much lower space
density (in addition, the comoving volume in the local universe
is smaller). Thus, for the same Compton-$y$ parameter, the
Poisson contribution to angular CMB fluctuations will be
negligible compared to that from SZ clusters. We have verified this by
direct numerical calculation.

We thus turn to the CMB fluctuations induced by clustering of
high-$z$ SN. We suppose for simplicity that stars form
only in halos where atomic cooling can operate, $T_{vir} > 10^{4}$K,
where some constant fraction $f_{*}$ of the baryonic mass fragments
to form stars. Both of these are normalized
to produce the same total fraction of baryons processed in VMSs
$f_{*,global}= 2.5 \times 10^{-4} (N_{\gamma}/25)$. We use
Press-Schechter theory to calculate the abundance of
halos. A hot bubble around each source has a
total flux $S \propto E_{SN}$ as given by equation
(\ref{eqn:SZ_flux}), and lasts for a Compton cooling
time $t_{C}$. The size of the hot bubble is given by equation
(\ref{eqn:size_bubble}). The finite bubble size damps the power
spectrum on scales below
the bubble size. For simplicity we shall assume $y_{l} = y_{o} {\rm
exp}[-(l/l_{c})^{2}]$, where $y_{l}$ is the Fourier transform of the $y$
profile of the bubble, $l_{c}=\pi/\theta_{c}$ and $\theta_{c}$ is the
angular size of the bubble when most of the Compton cooling takes
place. If the SF efficiency is independent of
halo mass then $y_{o} = R M_{halo}$, where the normalization constant
$R \propto \epsilon f_{*}$ is determined from the condition that:
\begin{equation}
     \bar{y}=\int dz (dV/dz d\Omega)
     \int_{M_{min}}^{\infty} dM (dn/dM) y_{o}(M,z),
\end{equation}
subject of course to the condition that $f_{coll}>f_{VMS}$ and
$\epsilon f_{*} < 1$.

In reionized regions, gas accretion is suppressed in halos with
$T_{vir} < T_{min} \approx 2.5
\times 10^{5}$K (or $v_{\rm cir} \sim 50 {\rm km \, s^{-1}}$,
\scite{thoul_weinberg}); lower-mass halos are thus unlikely to be able
to form stars.  This boosts the clustering bias of SF
systems as reionization proceeds, which increases
the strength of SZ anisotropies. To keep our analysis general,
we conservatively only require $T_{vir} >
10^{4}$K, but then show how increasing the Jeans mass would boost the
clustering bias thereby enhancing CMB fluctuations.

\begin{figure}
\psfig{file=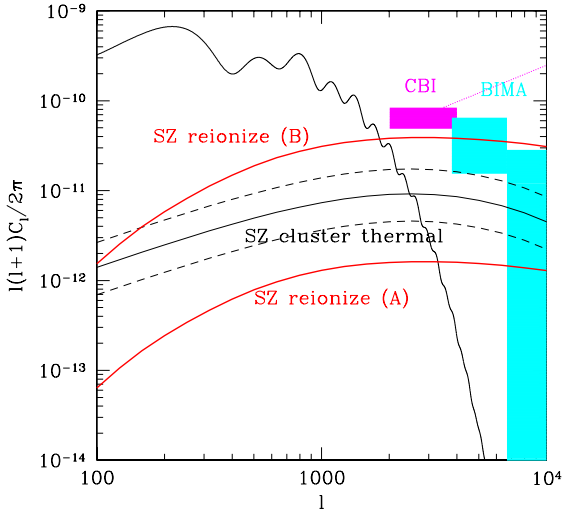,width=80mm}
\caption{The small-scale power spectrum of the CMB.  Our
standard case A for thermal SZ from reionization
assumes $y=3.6 \times 10^{-6}$ and clustering bias associated with
$T_{vir}=2 \times 10^{4}$K halos, while a maximal case B
assumes $y=10^{-5}$, and clustering bias associated
with $T_{vir}=2 \times 10^{4}$K halos. A
minimal case (not shown) which assumes a lower $y$ or lower
clustering bias would be down at the $10^{-14}-10^{-13}$ level. Also
shown are the power spectra due to physics at the surface of last
scatter, and the thermal-SZ effect for clusters for $\sigma_{8}=0.84
\pm 0.08 \,(2\sigma)$ (dashed lines are curves for $\sigma_{8}=0.76,0.92$).}
\label{fig:cls}
\end{figure}

The Compton $y$ power spectrum due to clustering of sources is given by:
\begin{eqnarray} 
C_{l}(y)= \int dz \frac{dV}{dz d\Omega}
P(k=\frac{l}{d_{M}(z)}) \\ \nonumber
\times \left[
\int_{M_{min}}^{\infty} dM \frac{dn}{dM} b(M,z) y_{l}(M,z) \right]^{2}
\end{eqnarray}
where $P(k)$ is the linear power spectrum, $d_M=d_{A}(1+z)$ is the
comoving angular diameter distance, and $b(M,z)$ is the linear bias
factor \cite{mo_white}. We have used the Limber approximation
$k=l/d_{M}$ which is valid for small angles. Note that $C_{l}(\Delta
T/T)=4 C_{l}(y)$in the RJ limit. The results are shown in
Figure \ref{fig:cls} for two cases: (A) a standard ``best-estimate'' case with $y=3.6 \times
10^{-6}$ and clustering bias associated with $T_{vir} > 2
\times 10^{4}$K halos; and (B) a maximal case with $y=10^{-5}$
(consistent with the current uncertainty in $\tau_{e}$,
$C_{II}$, and $f_{\rm esc}$), the largest value allowed by the COBE
constraint $y < 1.5 \times 10^{-5}$), and clustering bias
associated with $T_{vir} > 10^{5}$ K halos. Also shown are the
cluster-induced power spectra for $\sigma_{8}=0.84 \pm 0.08$
(2$\sigma$), computed as in \scite{Coo00}. Although the ``best-estimate'' reionization signal lies
below the cluster signal, with current uncertainties they could plausibly be
comparable. The shape of the power spectra are fairly well
constrained, but their amplitude is uncertain by $\sim 1-2$
orders of magnitude, as we discuss below.

\begin{figure}
\psfig{file=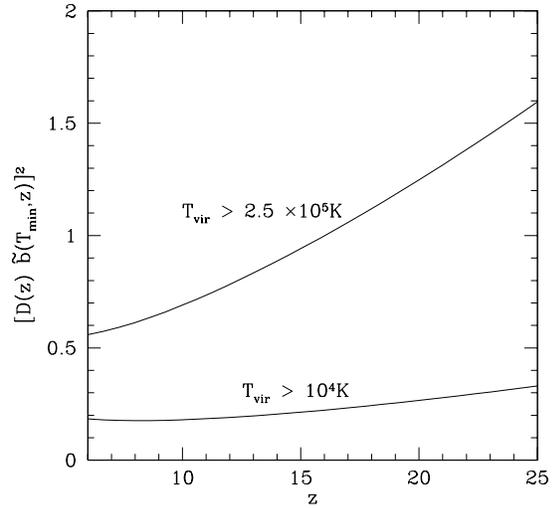,width=80mm}
\caption{The evolution of the normalization factor for the halo power
     spectrum, $[D(z) \tilde{b}(M,z)]^{2}$, where $D(z)$ is the growth factor
     and $\tilde{b}(M,z)$ is the mass-weighted bias from equation
     \ref{eqn:mass_weight_bias}. The halo power spectrum for halos where
     $T_{vir} > 10^{4}$K (which can cool by atomic cooling) hardly evolves
     with redshift; the increase in bias cancels the decrease in the growth
     factor. In addition, as reionization proceeds, halos with $T_{vir} <
     2.5 \times 10^{5}$K are unable to accrete gas; hence the clustering
     bias of SF halos will increase further. The
     thermal-SZ power spectra in 
     Fig 1 scale directly with this factor.}
\label{fig:bias}
\end{figure}

Roughly speaking, the CMB power spectrum is $C_{l}
\approx \bar{y}^{2} w_{l}$, where $\bar{y}$ is the
mean Compton $y$ parameter from equation \ref{eqn:y_parameter}, and
$w_{l} \propto l^n$ (if $P(k) \propto k^n$) is the flux-weighted
halo angular power spectrum. The flatness of $w_{l} l^{2}$ at
high $l$ is because $P(k) \propto k^{-2}$ at these wavenumber. 
For halos with $T_{\rm vir} > 10^{4}$K, the rapid increase in
bias tend to cancel the decrease in the growth factor at high
$z$, and the halo correlation function and power
spectrum $b(M_{1}) b(M_{2}) D(z)^{2} P(k)$ ($D(z)$ is the
linear-theory growth factor) do not evolve strongly
with redshift. We see this in Fig. \ref{fig:bias},
where we plot $[\tilde{b(M(T_{c}),z)}D(z)]^{2}$, and
$\tilde{b}(M(T_{c}),z)$ is the mass-weighted bias,
\begin{equation}
     \tilde{b}(M_{c},z)= \int_{M_{c}}^{\infty}dM \frac{dN}{dM}M
     b(M,z) \Bigg\slash \int_{M_{c}}^{\infty}dM \frac{dN}{dM}M,
\label{eqn:mass_weight_bias}
\end{equation}
which corresponds to the flux-weighted bias since we assume $S
\propto M$. This is likely a minimal estimate of the bias since the SF
efficiency (and hence the thermal SZ flux) is likely to increase with
the depth of the potential well. As reionization proceeds, the actual
bias interpolates between the two curves, since accretion is
suppressed in halos forming in reionized regions with $T_{vir} < 2.5
\times 10^{5}$K; it approaches the upper curve as $Q_{II}
\rightarrow 1$. Since we are probing scales on order of or
smaller than the halo correlation length, $r_{o} \sim {\rm
few} \,$Mpc comoving, it is reasonable to expect projected halo
density (and hence flux) enhancements of order $l (l+1)
w_{l}/(2\pi) \sim {\rm few}$.

Overall, our primary uncertainties in the predicted amplitude
are due to uncertainties in the mean $y$ parameter that arise
from the uncertainties in $\tau_{e}$, $C_{II}$, and $f_{\rm
esc}$ discussed above.  There is then an additional uncertainty
of $\sim$few introduced by the range of halo bias factors
illustrated in Fig. \ref{fig:bias}.

\section{Conclusions}

We have pointed out that WMAP's large electron-scattering optical
depth $\tau_{e}$ implies that SZ fluctuations from high-$z$ SF
could be considerable. As an interesting secondary result,
we derive a relation between $\tau_{e}$ and $N_{\gamma}$, the number
of ionizing photons emitted per baryon. We use this to calibrate
the amount of SN activity, and thereby obtain the expected
Compton-$y$ distortion, $y \sim {\rm few} \times 10^{-6}$.
Fluctuations in the Compton-$y$ parameter could be
detectable and may well account for the the small-scale
CMB-fluctuation excess at small angular scales. If so,
small-scale CMB measurements are {\it not} a reliable
independent measure of $\sigma_{8}$. If the small-scale CMB
anisotropies are due to clusters alone, they will be resolved by
forthcoming high-sensitivity and high-resolution SZ surveys.  On
the other hand, if high-$z$ SF contributes significantly, there
will be a substantial unresolved component, since the extremely faint
flux from individual halos is undetectable. A large amount of high
redshift SN activity also produces X-rays \cite{oh2001}, with
interesting consequences for reionization.

If a high-$z$ origin of the observed small-scale CMB
fluctuations is confirmed, CMB maps may then be used to study
the topology of reionization, perhaps by cross-correlating
with future 21cm tomographic maps of neutral hydrogen at high
$z$ \cite{tozzietal}.  Here we have focused exclusively on
thermal-SZ fluctuations, which induce a Compton-$y$ distortion
to the CMB frequency spectrum and can thus be distinguished from
``genuine'' temperature fluctuations with multifrequency CMB
measurements. However, high-$z$ SF may also induce temperature
fluctuations by scattering from reionized regions with coherent
large-scale peculiar velocities, as we detail in an accompanying
paper (Cooray et al., in preparation).

Given the uncertainties in high-$z$ SF discussed above,
we can make predictions for small-scale $y$ fluctuations with
roughly an order-of-magnitude level of uncertainty in the
CMB-fluctuation amplitude, and thus cannot at this point
conclusively attribute observed small-scale CMB-fluctuation
excesses to high-$z$ star formation.  Nonetheless, this
interpretation of the excess is certainly plausible.  If it is
correct, then the CMB experimentalists have achieved a
remarkable triumph: not only have they fulfilled a decade-old
quest to measure cosmological parameters with exquisite and
unprecedented precision, they have detected signatures of the
very first generation of star formation not just once, but twice.

\section*{Acknowledgments}
This work was supported by NASA NAG5-9821 and DoE
DE-FG03-92-ER40701.

\end{document}